# A Sequence of Generalizations of Cartan's Conservation of Torsion Theorem

C. C. Briggs
*Center for Academic Computing, Penn State University, University Park, PA 16802*
Monday, August 9, 1999

**Abstract.** A sequence of generalizations of Cartan's conservation of torsion theorem is given for $n$-dimensional differentiable manifolds having a general linear connection.
PACS numbers: 02.40.-k, 04.20.Fy

This paper presents a sequence of generalizations of Cartan's conservation of torsion theorem (also called "the 1$^{st}$ Bianchi identity") for $n$-dimensional differentiable manifolds having a general linear connection.

Cartan's "conservation of torsion theorem"[1] (from the French "théorème de conservation de la torsion"[2]) for such a manifold $M$ is given by[3-5]

$$\frac{1}{2} R_{[bcd]}{}^a = \nabla_{[b} S_{cd]}{}^a + 2 S_{[bc}{}^e S_{|e|d]}{}^a \qquad (1)$$
$$= \nabla_{[b} S_{cd]}{}^a - 2 S_{[bc}{}^e S_{d]e}{}^a,$$

where $R_{abc}{}^d$ is the Riemann-Christoffel curvature tensor and $S_{ab}{}^c$ the torsion tensor of $M$ as given by[6]

$$R_{abc}{}^d = 2 (\partial_{[a} \Gamma_{b]}{}^d{}_c + \Gamma_{[a|e|}{}^d \Gamma_{b]}{}^e{}_c + \Omega_a{}^e{}_b \Gamma_e{}^d{}_c) \qquad (2)$$

and[7]

$$S_{ab}{}^c = \Gamma_{[a\ b]}{}^c + \Omega_a{}^c{}_b, \qquad (3)$$

respectively, where $\Gamma_a{}^c{}_b$ is the connection coefficient and $\Omega_a{}^c{}_b$ the object of anholonomity.

The 1$^{st}$ ordinary exterior differentials[8] of the basis tangent vectors $\mathsf{e}_a$ of $M$ are given by[9-11]

$$\mathsf{d}\,\mathsf{e}_a = \mathsf{e}_b\,\omega_a{}^b, \qquad (4)$$

the contractions of which with the basis 1-forms $\omega^b$ of $M$ are given by

$$\langle \omega^b, \mathsf{d}\,\mathsf{e}_a \rangle = \omega_a{}^b \qquad (5)$$

and in view of which the 1$^{st}$ absolute exterior differentials[12-13] of $\mathsf{e}_a$ are given by

$$\mathsf{D}\,\mathsf{e}_a = \mathsf{d}\,\mathsf{e}_a - \mathsf{e}_b\,\omega_a{}^b \qquad (6)$$
$$= 0,$$

where $\omega_a{}^b$ is the connection 1-form of $M$ and where the contractions of $\omega^b$ with $\mathsf{e}_a$ are given by

$$\langle \omega^b, \mathsf{e}_a \rangle = \delta_a^b, \qquad (7)$$

where $\delta_a^b$ is the Kronecker delta.

The 2$^{nd}$ ordinary exterior differentials of $\mathsf{e}_a$ are given by[14-23]

$$\mathsf{d}^2\,\mathsf{e}_a = \mathsf{d}\,\mathsf{d}\,\mathsf{e}_a \qquad (8)$$
$$= \mathsf{d}\,\mathsf{e}_b\,\omega_a{}^b$$
$$= (\mathsf{d}\,\mathsf{e}_b) \wedge \omega_a{}^b + \mathsf{e}_b\,\mathsf{d}\,\omega_a{}^b$$
$$= \mathsf{e}_c\,\omega_b{}^c \wedge \omega_a{}^b + \mathsf{e}_b\,\mathsf{d}\,\omega_a{}^b$$
$$= \mathsf{e}_b\,(\mathsf{d}\,\omega_a{}^b + \omega_c{}^b \wedge \omega_a{}^c)$$
$$= \mathsf{e}_b\,\Omega_a{}^b,$$

the contractions of which with $\omega^b$ are given by

$$\langle \omega^b, \mathsf{d}^2\,\mathsf{e}_a \rangle = \Omega_a{}^b \qquad (9)$$

and where

$$\Omega_a{}^b = \mathsf{d}\,\omega_a{}^b + \omega_c{}^b \wedge \omega_a{}^c \qquad (10)$$
$$= \mathsf{D}\,\omega_a{}^b + \omega_a{}^c \wedge \omega_c{}^b$$
$$= \mathsf{d}\,\omega_a{}^b - \omega_a{}^c \wedge \omega_c{}^b$$
$$= \mathsf{D}\,\omega_a{}^b - \omega_c{}^b \wedge \omega_a{}^c$$
$$= \frac{1}{2} R_{cda}{}^b\,\omega^c \wedge \omega^d,$$

where $\Omega_a{}^b$ is the curvature 2-form of $M$.

The 3$^{rd}$ ordinary exterior differentials of $\mathsf{e}_a$ are given by

$$\mathsf{d}^3\,\mathsf{e}_a = \mathsf{d}\,\mathsf{d}^2\,\mathsf{e}_a \qquad (11)$$
$$= \mathsf{d}\,\mathsf{e}_b\,\Omega_a{}^b$$
$$= (\mathsf{d}\,\mathsf{e}_b) \wedge \Omega_a{}^b + \mathsf{e}_b\,\mathsf{d}\,\Omega_a{}^b$$
$$= (\mathsf{e}_c\,\omega_b{}^c) \wedge \Omega_a{}^b + \mathsf{e}_b\,\mathsf{d}\,\Omega_a{}^b$$
$$= \mathsf{e}_c\,\Omega_b{}^c \wedge \omega_a{}^b$$

using Bianchi's identity for $\Omega_a{}^b$,[24] i.e.,

$$\mathsf{D}\,\Omega_a{}^b = \mathsf{d}\,\Omega_a{}^b - \omega_a{}^c \wedge \Omega_c{}^b + \omega_c{}^b \wedge \Omega_a{}^c \qquad (12)$$
$$= 0,$$

as well as by

$$d^3 e_a = d^2 d e_a \tag{13}$$
$$= d^2 e_b \, \omega_a{}^b$$
$$= (d^2 e_b) \wedge \omega_a{}^b + e_b \, d^2 \omega_a{}^b$$
$$= (e_c \, \Omega_b{}^c) \wedge \omega_a{}^b + 0$$
$$= e_c \, \Omega_b{}^c \wedge \omega_a{}^b$$

using Poincaré's theorem for scalar-valued exterior differential forms,[25-26] i.e.,

$$d^2 \alpha = 0, \tag{14}$$

where $\alpha$ is an arbitrary scalar-valued exterior differential form.

The $4^{th}$ ordinary exterior differentials of $e_a$ are given by

$$d^4 e_a = d^2 d^2 e_a \tag{15}$$
$$= d^2 e_b \, \Omega_a{}^b$$
$$= (d^2 e_b) \wedge \Omega_a{}^b + e_b \, d^2 \Omega_a{}^b$$
$$= (e_c \, \Omega_b{}^c) \wedge \Omega_a{}^b + 0$$
$$= e_c \, \Omega_b{}^c \wedge \Omega_a{}^b.$$

In general, the $p^{th}$ ordinary exterior differentials of $e_a$ for $p > 0$ are given (cf. Flanders[27]) by

$$d^p e_a = \begin{cases} e_{i_{(p+1)/2}} \, \omega_a{}^{i_1} \wedge \Omega_{i_1}{}^{i_2} \wedge \Omega_{i_2}{}^{i_3} \wedge \ldots \wedge \Omega_{i_{(p-5)/2}}{}^{i_{(p-3)/2}} \wedge \Omega_{i_{(p-3)/2}}{}^{i_{(p-1)/2}} \wedge \Omega_{i_{(p-1)/2}}{}^{i_{(p+1)/2}}, & \text{if } p \text{ is odd} \\ e_{i_{p/2}} \, \Omega_a{}^{i_1} \wedge \Omega_{i_1}{}^{i_2} \wedge \Omega_{i_2}{}^{i_3} \wedge \ldots \wedge \Omega_{i_{(p-6)/2}}{}^{i_{(p-4)/2}} \wedge \Omega_{i_{(p-4)/2}}{}^{i_{(p-2)/2}} \wedge \Omega_{i_{(p-2)/2}}{}^{i_{p/2}}, & \text{if } p \text{ is even} \end{cases}, \tag{16}$$

the contractions of which with $\omega^a$ are given by

$$\langle \omega^a, d^p e_a \rangle = \begin{cases} \omega_{i_{(p+1)/2}}{}^{i_1} \wedge \Omega_{i_1}{}^{i_2} \wedge \Omega_{i_2}{}^{i_3} \wedge \ldots \wedge \Omega_{i_{(p-5)/2}}{}^{i_{(p-3)/2}} \wedge \Omega_{i_{(p-3)/2}}{}^{i_{(p-1)/2}} \wedge \Omega_{i_{(p-1)/2}}{}^{i_{(p+1)/2}}, & \text{if } p \text{ is odd} \\ \Omega_{i_{p/2}}{}^{i_1} \wedge \Omega_{i_1}{}^{i_2} \wedge \Omega_{i_2}{}^{i_3} \wedge \ldots \wedge \Omega_{i_{(p-6)/2}}{}^{i_{(p-4)/2}} \wedge \Omega_{i_{(p-4)/2}}{}^{i_{(p-2)/2}} \wedge \Omega_{i_{(p-2)/2}}{}^{i_{p/2}}, & \text{if } p \text{ is even} \end{cases}. \tag{17}$$

"The soldering form"[28-30] (also called "the Cartan-Maurer form,"[31] "the canonical form,"[32-33] and "the displacement vector"[34-35]) $d\,P$ of $M$ is the $1^{st}$ ordinary exterior differential of the vector-valued 0-form $P$ and is given in terms of $e_a$ and $\omega^a$ by

$$d\,P = e_a \otimes \omega^a = e_a \, \omega^a = e_a \, \delta_b^a \, \omega^b, \tag{18}$$

the coefficients of which[36-37] are equal to $\delta_b^a$, i.e., to the Kronecker delta.

The $1^{st}$ ordinary exterior differential of $d\,P$ is the $2^{nd}$ ordinary exterior differential of $P$, i.e., $d^2 P$, and is given by[38-39]

$$d^2 P = d\, d\, P \tag{19}$$
$$= d\,(e_a \, \omega^a)$$
$$= (d\, e_a) \wedge \omega^a + e_a \, d \omega^a$$
$$= (e_b \, \omega_a{}^b) \wedge \omega^a + e_a \, d \omega^a$$
$$= e_a \,(d \omega^a + \omega_b{}^a \wedge \omega^b)$$
$$= e_a \, D \omega^a$$
$$= e_a \, \Omega^a,$$

where $\Omega^a$ is the torsion 2-form defined by

$$\Omega^a = D \omega^a \tag{20}$$
$$= d \omega^a + \omega_b{}^a \wedge \omega^b.$$

The $1^{st}$ ordinary exterior differential of $\Omega^a$ is given by

$$d \Omega^a = d\,(d \omega^a + \omega_b{}^a \wedge \omega^b) \tag{21}$$
$$= d^2 \omega^a + (d \omega_b{}^a) \wedge \omega^b - \omega_b{}^a \wedge d \omega^b$$
$$= 0 + (\Omega_b{}^a + \omega_b{}^c \wedge \omega_c{}^a) \wedge \omega^b - \omega_b{}^a \wedge (\Omega^b - \omega_c{}^b \wedge \omega^c)$$
$$= \Omega_b{}^a \wedge \omega^b + \omega_b{}^c \wedge \omega_c{}^a \wedge \omega^b - \omega_b{}^a \wedge \Omega^b + \omega_b{}^a \wedge \omega_c{}^b \wedge \omega^c$$
$$= \Omega_b{}^a \wedge \omega^b - \omega_b{}^a \wedge \Omega^b,$$

in view of which Cartan's conservation of torsion theorem is given also by

$$D \Omega^a = d \Omega^a + \omega_b{}^a \wedge \Omega^b \tag{22}$$
$$= \Omega_b{}^a \wedge \omega^b.$$

The $3^{rd}$ ordinary exterior differential of $P$ is given by

$$d^3 P = d\, d^2 P \tag{23}$$
$$= d\, e_a \, \Omega^a$$
$$= (d\, e_a) \wedge \Omega^a + e_a \, d \Omega^a$$
$$= e_a \, d \Omega^a + (e_b \, \omega_a{}^b) \wedge \Omega^a$$
$$= e_a \,(d \Omega^a + \omega_b{}^a \wedge \Omega^b)$$
$$= e_a \, D \Omega^a$$
$$= d^2 \, d\, P$$
$$= d^2 \, e_a \, \omega^a$$
$$= (d^2 e_a) \wedge \omega^a + e_a \, d^2 \omega^a$$
$$= (e_b \, \Omega_a{}^b) \wedge \omega^a + 0$$
$$= e_a \, \Omega_b{}^a \wedge \omega^b.$$

The $4^{th}$ ordinary exterior differential of $P$ is given by

$$d^4 P = d^2 \, d^2 P \tag{24}$$
$$= d^2 e_a \, \Omega^a$$
$$= (d^2 e_a) \wedge \Omega^a + e_a \, d^2 \Omega^a$$
$$= (e_b \, \Omega_a{}^b) \wedge \Omega^a + 0$$
$$= e_a \, \Omega_b{}^a \wedge \Omega^b.$$

For $1 \le p \le 10$, the vector-valued $p$-forms $\mathsf{d}^p \mathsf{P}$ are given by

$$\mathsf{d}\,\mathsf{P} = \mathsf{e}_a\,\omega^a, \tag{25}$$

$$\mathsf{d}^2\,\mathsf{P} = \mathsf{e}_a\,\mathsf{D}\,\omega^a \tag{26}$$
$$= \mathsf{e}_a\,\Omega^a,$$

$$\mathsf{d}^3\,\mathsf{P} = \mathsf{e}_a\,\mathsf{D}^2\,\omega^a \tag{27}$$
$$= \mathsf{e}_a\,\mathsf{D}\,\Omega^a$$
$$= \mathsf{e}_a\,\Omega_b{}^a \wedge \omega^b,$$

$$\mathsf{d}^4\,\mathsf{P} = \mathsf{e}_a\,\mathsf{D}^3\,\omega^a \tag{28}$$
$$= \mathsf{e}_a\,\mathsf{D}^2\,\Omega^a$$
$$= \mathsf{e}_a\,\Omega_b{}^a \wedge \mathsf{D}\,\omega^b$$
$$= \mathsf{e}_a\,\Omega_b{}^a \wedge \Omega^b,$$

$$\mathsf{d}^5\,\mathsf{P} = \mathsf{e}_a\,\mathsf{D}^4\,\omega^a \tag{29}$$
$$= \mathsf{e}_a\,\mathsf{D}^3\,\Omega^a$$
$$= \mathsf{e}_a\,\Omega_b{}^a \wedge \mathsf{D}^2\,\omega^b$$
$$= \mathsf{e}_a\,\Omega_b{}^a \wedge \mathsf{D}\,\Omega^b$$
$$= \mathsf{e}_a\,\Omega_b{}^a \wedge \Omega_c{}^b \wedge \omega^c,$$

$$\mathsf{d}^6\,\mathsf{P} = \mathsf{e}_a\,\mathsf{D}^5\,\omega^a \tag{30}$$
$$= \mathsf{e}_a\,\mathsf{D}^4\,\Omega^a$$
$$= \mathsf{e}_a\,\Omega_b{}^a \wedge \mathsf{D}^3\,\omega^b$$
$$= \mathsf{e}_a\,\Omega_b{}^a \wedge \mathsf{D}^2\,\Omega^b$$
$$= \mathsf{e}_a\,\Omega_b{}^a \wedge \Omega_c{}^b \wedge \mathsf{D}\,\omega^c$$
$$= \mathsf{e}_a\,\Omega_b{}^a \wedge \Omega_c{}^b \wedge \Omega^c,$$

$$\mathsf{d}^7\,\mathsf{P} = \mathsf{e}_a\,\mathsf{D}^6\,\omega^a \tag{31}$$
$$= \mathsf{e}_a\,\mathsf{D}^5\,\Omega^a$$
$$= \mathsf{e}_a\,\Omega_b{}^a \wedge \mathsf{D}^4\,\omega^b$$
$$= \mathsf{e}_a\,\Omega_b{}^a \wedge \mathsf{D}^3\,\Omega^b$$
$$= \mathsf{e}_a\,\Omega_b{}^a \wedge \Omega_c{}^b \wedge \mathsf{D}^2\,\omega^c$$
$$= \mathsf{e}_a\,\Omega_b{}^a \wedge \Omega_c{}^b \wedge \mathsf{D}\,\Omega^c$$
$$= \mathsf{e}_a\,\Omega_b{}^a \wedge \Omega_c{}^b \wedge \Omega_d{}^c \wedge \omega^d,$$

$$\mathsf{d}^8\,\mathsf{P} = \mathsf{e}_a\,\mathsf{D}^7\,\omega^a \tag{32}$$
$$= \mathsf{e}_a\,\mathsf{D}^6\,\Omega^a$$
$$= \mathsf{e}_a\,\Omega_b{}^a \wedge \mathsf{D}^5\,\omega^b$$
$$= \mathsf{e}_a\,\Omega_b{}^a \wedge \mathsf{D}^4\,\Omega^b$$
$$= \mathsf{e}_a\,\Omega_b{}^a \wedge \Omega_c{}^b \wedge \mathsf{D}^3\,\omega^c$$
$$= \mathsf{e}_a\,\Omega_b{}^a \wedge \Omega_c{}^b \wedge \mathsf{D}^2\,\Omega^c$$
$$= \mathsf{e}_a\,\Omega_b{}^a \wedge \Omega_c{}^b \wedge \Omega_d{}^c \wedge \mathsf{D}\,\omega^d$$
$$= \mathsf{e}_a\,\Omega_b{}^a \wedge \Omega_c{}^b \wedge \Omega_d{}^c \wedge \Omega^d,$$

$$\mathsf{d}^9\,\mathsf{P} = \mathsf{e}_a\,\mathsf{D}^8\,\omega^a \tag{33}$$
$$= \mathsf{e}_a\,\mathsf{D}^7\,\Omega^a$$
$$= \mathsf{e}_a\,\Omega_b{}^a \wedge \mathsf{D}^6\,\omega^b$$
$$= \mathsf{e}_a\,\Omega_b{}^a \wedge \mathsf{D}^5\,\Omega^b$$
$$= \mathsf{e}_a\,\Omega_b{}^a \wedge \Omega_c{}^b \wedge \mathsf{D}^4\,\omega^c$$
$$= \mathsf{e}_a\,\Omega_b{}^a \wedge \Omega_c{}^b \wedge \mathsf{D}^3\,\Omega^c$$
$$= \mathsf{e}_a\,\Omega_b{}^a \wedge \Omega_c{}^b \wedge \Omega_d{}^c \wedge \mathsf{D}^2\,\omega^d$$
$$= \mathsf{e}_a\,\Omega_b{}^a \wedge \Omega_c{}^b \wedge \Omega_d{}^c \wedge \mathsf{D}\,\Omega^d$$
$$= \mathsf{e}_a\,\Omega_b{}^a \wedge \Omega_c{}^b \wedge \Omega_d{}^c \wedge \Omega_e{}^d \wedge \omega^e,$$

$$\mathsf{d}^{10}\,\mathsf{P} = \mathsf{e}_a\,\mathsf{D}^9\,\omega^a \tag{34}$$
$$= \mathsf{e}_a\,\mathsf{D}^8\,\Omega^a$$
$$= \mathsf{e}_a\,\Omega_b{}^a \wedge \mathsf{D}^7\,\omega^b$$
$$= \mathsf{e}_a\,\Omega_b{}^a \wedge \mathsf{D}^6\,\Omega^b$$
$$= \mathsf{e}_a\,\Omega_b{}^a \wedge \Omega_c{}^b \wedge \mathsf{D}^5\,\omega^c$$
$$= \mathsf{e}_a\,\Omega_b{}^a \wedge \Omega_c{}^b \wedge \mathsf{D}^4\,\Omega^c$$
$$= \mathsf{e}_a\,\Omega_b{}^a \wedge \Omega_c{}^b \wedge \Omega_d{}^c \wedge \mathsf{D}^3\,\omega^d$$
$$= \mathsf{e}_a\,\Omega_b{}^a \wedge \Omega_c{}^b \wedge \Omega_d{}^c \wedge \mathsf{D}^2\,\Omega^d$$
$$= \mathsf{e}_a\,\Omega_b{}^a \wedge \Omega_c{}^b \wedge \Omega_d{}^c \wedge \Omega_e{}^d \wedge \mathsf{D}\,\omega^e$$
$$= \mathsf{e}_a\,\Omega_b{}^a \wedge \Omega_c{}^b \wedge \Omega_d{}^c \wedge \Omega_e{}^d \wedge \Omega^e.$$

For suitably high values of $p$, the vector-valued $p$-forms $\mathsf{d}^p \mathsf{P}$ are given by

$$\mathsf{d}^p\,\mathsf{P} = \mathsf{e}_a\,\mathsf{D}^{p-1}\,\omega^a \tag{35}$$
$$= \mathsf{e}_a\,\mathsf{D}^{p-2}\,\Omega^a$$
$$= \mathsf{e}_a\,\Omega_b{}^a \wedge \mathsf{D}^{p-3}\,\omega^b$$
$$= \mathsf{e}_a\,\Omega_b{}^a \wedge \mathsf{D}^{p-4}\,\Omega^b$$
$$= \mathsf{e}_a\,\Omega_b{}^a \wedge \Omega_c{}^b \wedge \mathsf{D}^{p-5}\,\omega^c$$
$$= \mathsf{e}_a\,\Omega_b{}^a \wedge \Omega_c{}^b \wedge \mathsf{D}^{p-6}\,\Omega^c$$
$$= \mathsf{e}_a\,\Omega_b{}^a \wedge \Omega_c{}^b \wedge \Omega_d{}^c \wedge \mathsf{D}^{p-7}\,\omega^d$$
$$= \mathsf{e}_a\,\Omega_b{}^a \wedge \Omega_c{}^b \wedge \Omega_d{}^c \wedge \mathsf{D}^{p-8}\,\Omega^d$$
$$= \mathsf{e}_a\,\Omega_b{}^a \wedge \Omega_c{}^b \wedge \Omega_d{}^c \wedge \Omega_e{}^d \wedge \mathsf{D}^{p-9}\,\omega^e$$
$$= \mathsf{e}_a\,\Omega_b{}^a \wedge \Omega_c{}^b \wedge \Omega_d{}^c \wedge \Omega_e{}^d \wedge \mathsf{D}^{p-10}\,\Omega^e,$$

and so on.

In general, for $p > 2$, the vector-valued $p$-forms $\mathsf{d}^p \mathsf{P}$ are given (cf. Flanders[40]) by

$$\begin{aligned}
\mathsf{d}^p\,\mathsf{P} &= \mathsf{e}_a\,\mathsf{D}^{p-1}\,\omega^a \\
&= \mathsf{e}_a\,\mathsf{D}^{p-2}\,\Omega^a \\
&\phantom{=}\ \vdots \\
&= \begin{cases} \mathsf{e}_a\,\Omega_{i_1}{}^a \wedge \Omega_{i_2}{}^{i_1} \wedge \Omega_{i_3}{}^{i_2} \wedge \ldots \wedge \Omega_{i_{(p-3)/2}}{}^{i_{(p-5)/2}} \wedge \Omega_{i_{(p-1)/2}}{}^{i_{(p-3)/2}} \wedge \omega^{i_{(p-1)/2}}, & \text{if } p \text{ is odd} \\ \mathsf{e}_a\,\Omega_{i_1}{}^a \wedge \Omega_{i_2}{}^{i_1} \wedge \Omega_{i_3}{}^{i_2} \wedge \ldots \wedge \Omega_{i_{(p-4)/2}}{}^{i_{(p-6)/2}} \wedge \Omega_{i_{(p-2)/2}}{}^{i_{(p-4)/2}} \wedge \Omega^{i_{(p-2)/2}}, & \text{if } p \text{ is even} \end{cases} \\
&= \begin{cases} \dfrac{1}{2^{(p-1)/2}}\,\mathsf{e}_a\,R_{[i_1 i_2 | j_1|}{}^a\,R_{i_3 i_4 | j_2|}{}^{j_1}\,R_{i_5 i_6 | j_3|}{}^{j_2} \ldots R_{i_{p-4} i_{p-3} | j_{(p-3)/2}|}{}^{j_{(p-5)/2}}\,R_{i_{p-2} i_{p-1} p]}{}^{j_{(p-3)/2}}\,\omega^{i_1} \wedge \omega^{i_2} \wedge \ldots \wedge \omega^{i_p}, & \text{if } p \text{ is odd} \\ \dfrac{1}{2^{(p-2)/2}}\,\mathsf{e}_a\,R_{[i_1 i_2 | j_1|}{}^a\,R_{i_3 i_4 | j_2|}{}^{j_1}\,R_{i_5 i_6 | j_3|}{}^{j_2} \ldots R_{i_{p-3} i_{p-2} | j_{(p-2)/2}|}{}^{j_{(p-4)/2}}\,S_{i_{p-1} i_p]}{}^{j_{(p-2)/2}}\,\omega^{i_1} \wedge \omega^{i_2} \wedge \ldots \wedge \omega^{i_p}, & \text{if } p \text{ is even} \end{cases}
\end{aligned} \tag{36}$$

Additional expressions for the vector-valued $p$-forms $\mathsf{d}^p \mathsf{P}$ and their coefficients for $1 \le p \le 10$ are given in the Appendix.

---

[40] Flanders, H., (1953), *op. cit.*, Eq. (9.4).

For $1 \leq p \leq 10$, the scalar-valued $p$-forms $\langle \omega^a, d^p P \rangle$ are given by

$$\langle \omega^a, d P \rangle = \omega^a \tag{37}$$

$$\langle \omega^a, d^2 P \rangle = D\, \omega^a \tag{38}$$
$$= \Omega^a,$$

$$\langle \omega^a, d^3 P \rangle = D^2 \omega^a \tag{39}$$
$$= D\, \Omega^a$$
$$= \Omega_b{}^a \wedge \omega^b,$$

$$\langle \omega^a, d^4 P \rangle = D^3 \omega^a \tag{40}$$
$$= D^2 \Omega^a$$
$$= \Omega_b{}^a \wedge D\, \omega^b$$
$$= \Omega_b{}^a \wedge \Omega^b,$$

$$\langle \omega^a, d^5 P \rangle = D^4 \omega^a \tag{41}$$
$$= D^3 \Omega^a$$
$$= \Omega_b{}^a \wedge D^2 \omega^b$$
$$= \Omega_b{}^a \wedge D\, \Omega^b$$
$$= \Omega_b{}^a \wedge \Omega_c{}^b \wedge \omega^c,$$

$$\langle \omega^a, d^6 P \rangle = D^5 \omega^a \tag{42}$$
$$= D^4 \Omega^a$$
$$= \Omega_b{}^a \wedge D^3 \omega^b$$
$$= \Omega_b{}^a \wedge D^2 \Omega^b$$
$$= \Omega_b{}^a \wedge \Omega_c{}^b \wedge D\, \omega^c$$
$$= \Omega_b{}^a \wedge \Omega_c{}^b \wedge \Omega^c,$$

$$\langle \omega^a, d^7 P \rangle = D^6 \omega^a \tag{43}$$
$$= D^5 \Omega^a$$
$$= \Omega_b{}^a \wedge D^4 \omega^b$$
$$= \Omega_b{}^a \wedge D^3 \Omega^b$$
$$= \Omega_b{}^a \wedge \Omega_c{}^b \wedge D^2 \omega^c$$
$$= \Omega_b{}^a \wedge \Omega_c{}^b \wedge D\, \Omega^c$$
$$= \Omega_b{}^a \wedge \Omega_c{}^b \wedge \Omega_d{}^c \wedge \omega^d,$$

$$\langle \omega^a, d^8 P \rangle = D^7 \omega^a \tag{44}$$
$$= D^6 \Omega^a$$
$$= \Omega_b{}^a \wedge D^5 \omega^b$$
$$= \Omega_b{}^a \wedge D^4 \Omega^b$$
$$= \Omega_b{}^a \wedge \Omega_c{}^b \wedge D^3 \omega^c$$
$$= \Omega_b{}^a \wedge \Omega_c{}^b \wedge D^2 \Omega^c$$
$$= \Omega_b{}^a \wedge \Omega_c{}^b \wedge \Omega_d{}^c \wedge D\, \omega^d$$
$$= \Omega_b{}^a \wedge \Omega_c{}^b \wedge \Omega_d{}^c \wedge \Omega^d,$$

$$\langle \omega^a, d^9 P \rangle = D^8 \omega^a \tag{45}$$
$$= D^7 \Omega^a$$
$$= \Omega_b{}^a \wedge D^6 \omega^b$$
$$= \Omega_b{}^a \wedge D^5 \Omega^b$$
$$= \Omega_b{}^a \wedge \Omega_c{}^b \wedge D^4 \omega^c$$
$$= \Omega_b{}^a \wedge \Omega_c{}^b \wedge D^3 \Omega^c$$
$$= \Omega_b{}^a \wedge \Omega_c{}^b \wedge \Omega_d{}^c \wedge D^2 \omega^d$$
$$= \Omega_b{}^a \wedge \Omega_c{}^b \wedge \Omega_d{}^c \wedge D\, \Omega^d$$
$$= \Omega_b{}^a \wedge \Omega_c{}^b \wedge \Omega_d{}^c \wedge \Omega_e{}^d \wedge \omega^e,$$

$$\langle \omega^a, d^{10} P \rangle = D^9 \omega^a \tag{46}$$
$$= D^8 \Omega^a$$
$$= \Omega_b{}^a \wedge D^7 \omega^b$$
$$= \Omega_b{}^a \wedge D^6 \Omega^b$$
$$= \Omega_b{}^a \wedge \Omega_c{}^b \wedge D^5 \omega^c$$
$$= \Omega_b{}^a \wedge \Omega_c{}^b \wedge D^4 \Omega^c$$
$$= \Omega_b{}^a \wedge \Omega_c{}^b \wedge \Omega_d{}^c \wedge D^3 \omega^d$$
$$= \Omega_b{}^a \wedge \Omega_c{}^b \wedge \Omega_d{}^c \wedge D^2 \Omega^d$$
$$= \Omega_b{}^a \wedge \Omega_c{}^b \wedge \Omega_d{}^c \wedge \Omega_e{}^d \wedge D\, \omega^e$$
$$= \Omega_b{}^a \wedge \Omega_c{}^b \wedge \Omega_d{}^c \wedge \Omega_e{}^d \wedge \Omega^e.$$

For suitably high values of $p$, the scalar-valued $p$-forms $\langle \omega^a, d^p P \rangle$ are given by

$$\langle \omega^a, d^p P \rangle = D^{p-1} \omega^a \tag{47}$$
$$= D^{p-2} \Omega^a$$
$$= \Omega_b{}^a \wedge D^{p-3} \omega^b$$
$$= \Omega_b{}^a \wedge D^{p-4} \Omega^b$$
$$= \Omega_b{}^a \wedge \Omega_c{}^b \wedge D^{p-5} \omega^c$$
$$= \Omega_b{}^a \wedge \Omega_c{}^b \wedge D^{p-6} \Omega^c$$
$$= \Omega_b{}^a \wedge \Omega_c{}^b \wedge \Omega_d{}^c \wedge D^{p-7} \omega^d$$
$$= \Omega_b{}^a \wedge \Omega_c{}^b \wedge \Omega_d{}^c \wedge D^{p-8} \Omega^d$$
$$= \Omega_b{}^a \wedge \Omega_c{}^b \wedge \Omega_d{}^c \wedge \Omega_e{}^d \wedge D^{p-9} \omega^e$$
$$= \Omega_b{}^a \wedge \Omega_c{}^b \wedge \Omega_d{}^c \wedge \Omega_e{}^d \wedge D^{p-10} \Omega^e,$$

and so on.

In general, for $p > 2$, the scalar-valued $p$-forms $\langle \omega^a, d^p P \rangle$ are given by

$$\langle \omega^a, d^p P \rangle = D^{p-1} \omega^a \tag{48}$$
$$= D^{p-2} \Omega^a$$
$$\quad \vdots$$
$$= \begin{cases} \Omega_{i_1}{}^a \wedge \Omega_{i_2}{}^{i_1} \wedge \Omega_{i_3}{}^{i_2} \wedge \ldots \wedge \Omega_{i_{(p-3)/2}}{}^{i_{(p-5)/2}} \wedge \Omega_{i_{(p-1)/2}}{}^{i_{(p-3)/2}} \wedge \omega^{i_{(p-1)/2}}, & \text{if } p \text{ is odd} \\ \Omega_{i_1}{}^a \wedge \Omega_{i_2}{}^{i_1} \wedge \Omega_{i_3}{}^{i_2} \wedge \ldots \wedge \Omega_{i_{(p-4)/2}}{}^{i_{(p-6)/2}} \wedge \Omega_{i_{(p-2)/2}}{}^{i_{(p-4)/2}} \wedge \Omega^{i_{(p-2)/2}}, & \text{if } p \text{ is even} \end{cases}$$

$$= \begin{cases} \dfrac{1}{2^{(p-1)/2}} R_{[i_1 i_2 | j_1|}{}^a R_{i_3 i_4 | j_2|}{}^{j_1} R_{i_5 i_6 | j_3|}{}^{j_2} \ldots R_{i_{p-4} i_{p-3} | j_{(p-3)/2}|}{}^{j_{(p-5)/2}} R_{i_{p-2} i_{p-1} i_p]}{}^{j_{(p-3)/2}} \omega^{i_1} \wedge \omega^{i_2} \wedge \ldots \wedge \omega^{i_p}, & \text{if } p \text{ is odd} \\ \dfrac{1}{2^{(p-2)/2}} R_{[i_1 i_2 | j_1|}{}^a R_{i_3 i_4 | j_2|}{}^{j_1} R_{i_5 i_6 | j_3|}{}^{j_2} \ldots R_{i_{p-3} i_{p-2} | j_{(p-2)/2}|}{}^{j_{(p-4)/2}} S_{i_{p-1} i_p]}{}^{j_{(p-2)/2}} \omega^{i_1} \wedge \omega^{i_2} \wedge \ldots \wedge \omega^{i_p}, & \text{if } p \text{ is even} \end{cases}.$$

Thus, a sequence of generalizations of Cartan's conservation of torsion theorem is given, for $p > 2$, by

$$D^{p-2} \Omega^a = \begin{cases} \Omega_{i_1}{}^a \wedge \Omega_{i_2}{}^{i_1} \wedge \Omega_{i_3}{}^{i_2} \wedge \ldots \wedge \Omega_{i_{(p-3)/2}}{}^{i_{(p-5)/2}} \wedge \Omega_{i_{(p-1)/2}}{}^{i_{(p-3)/2}} \wedge \omega^{i_{(p-1)/2}}, & \text{if } p \text{ is odd} \\ \Omega_{i_1}{}^a \wedge \Omega_{i_2}{}^{i_1} \wedge \Omega_{i_3}{}^{i_2} \wedge \ldots \wedge \Omega_{i_{(p-4)/2}}{}^{i_{(p-6)/2}} \wedge \Omega_{i_{(p-2)/2}}{}^{i_{(p-4)/2}} \wedge \Omega^{i_{(p-2)/2}}, & \text{if } p \text{ is even} \end{cases} \quad (49)$$

or equivalently, for $p > 0$, by

$$D^p \Omega^a = \begin{cases} \Omega_{i_1}{}^a \wedge \Omega_{i_2}{}^{i_1} \wedge \Omega_{i_3}{}^{i_2} \wedge \ldots \wedge \Omega_{i_{(p-1)/2}}{}^{i_{(p-3)/2}} \wedge \Omega_{i_{(p+1)/2}}{}^{i_{(p-1)/2}} \wedge \omega^{i_{(p+1)/2}}, & \text{if } p \text{ is odd} \\ \Omega_{i_1}{}^a \wedge \Omega_{i_2}{}^{i_1} \wedge \Omega_{i_3}{}^{i_2} \wedge \ldots \wedge \Omega_{i_{(p-2)/2}}{}^{i_{(p-4)/2}} \wedge \Omega_{i_{p/2}}{}^{i_{(p-2)/2}} \wedge \Omega^{i_{p/2}}, & \text{if } p \text{ is even} \end{cases} \quad (50)$$

## APPENDIX. ADDITIONAL EXPRESSIONS FOR THE VECTOR-VALUED $p$-FORMS $d^p P$ AND THEIR COEFFICIENTS FOR $1 \leq p \leq 10$

Expressions for the coefficients of the vector-valued $p$-forms $d^p P$ as given for $1 \leq p \leq 10$ by

$$d^1 P = e_a \, \omega^a \quad (51)$$
$$= e_a \, \delta_b^a \, \omega^b,$$
$$d^2 P = e_a \, \Omega^a \quad (52)$$
$$= e_a \, S_{bc}{}^a \, \omega^b \wedge \omega^c,$$
$$d^3 P = e_a \, \Omega_b{}^a \wedge \omega^b \quad (53)$$
$$= \tfrac{1}{2} e_a \, R_{[bcd]}{}^a \, \omega^b \wedge \omega^c \wedge \omega^d,$$
$$d^4 P = e_a \, \Omega_b{}^a \wedge \Omega^b \quad (54)$$
$$= \tfrac{1}{2} e_a \, R_{[bc|f|}{}^a \, S_{de]}{}^f \, \omega^b \wedge \omega^c \wedge \omega^d \wedge \omega^e,$$
$$d^5 P = e_a \, \Omega_b{}^a \wedge \Omega_c{}^b \wedge \omega^c \quad (55)$$
$$= \tfrac{1}{4} e_a \, R_{[bc|g|}{}^a \, R_{def]}{}^g \, \omega^b \wedge \omega^c \wedge \omega^d \wedge \omega^e \wedge \omega^f,$$
$$d^6 P = e_a \, \Omega_b{}^a \wedge \Omega_c{}^b \wedge \Omega^c \quad (56)$$
$$= \tfrac{1}{4} e_a \, R_{[bc|h|}{}^a \, R_{de|i|}{}^h \, S_{fg]}{}^i \, \omega^b \wedge \omega^c \wedge \omega^d \wedge \omega^e \wedge \omega^f \wedge \omega^g,$$
$$d^7 P = e_a \, \Omega_b{}^a \wedge \Omega_c{}^b \wedge \Omega_d{}^c \wedge \omega^d \quad (57)$$
$$= \tfrac{1}{8} e_a \, R_{[bc|i|}{}^a \, R_{de|j|}{}^i \, R_{fgh]}{}^j \, \omega^b \wedge \omega^c \wedge \omega^d \wedge \omega^e \wedge \omega^f \wedge \omega^g \wedge \omega^h,$$
$$d^8 P = e_a \, \Omega_b{}^a \wedge \Omega_c{}^b \wedge \Omega_d{}^c \wedge \Omega^d \quad (58)$$
$$= \tfrac{1}{8} e_a \, R_{[bc|j|}{}^a \, R_{de|k|}{}^j \, R_{fg|l|}{}^k \, S_{hi]}{}^l \, \omega^b \wedge \omega^c \wedge \omega^d \wedge \omega^e \wedge \omega^f \wedge \omega^g \wedge \omega^h \wedge \omega^i,$$
$$d^9 P = e_a \, \Omega_b{}^a \wedge \Omega_c{}^b \wedge \Omega_d{}^c \wedge \Omega_e{}^d \wedge \omega^e \quad (59)$$
$$= \tfrac{1}{16} e_a \, R_{[bc|k|}{}^a \, R_{de|l|}{}^k \, R_{fg|m|}{}^l \, R_{hij]}{}^m \, \omega^b \wedge \omega^c \wedge \omega^d \wedge \omega^e \wedge \omega^f \wedge \omega^g \wedge \omega^h \wedge \omega^i \wedge \omega^j,$$
$$d^{10} P = e_a \, \Omega_b{}^a \wedge \Omega_c{}^b \wedge \Omega_d{}^c \wedge \Omega_e{}^d \wedge \Omega^e \quad (60)$$
$$= \tfrac{1}{16} e_a \, R_{[bc|l|}{}^a \, R_{de|m|}{}^l \, R_{fg|n|}{}^m \, R_{hi|o|}{}^n \, S_{jk]}{}^o \, \omega^b \wedge \omega^c \wedge \omega^d \wedge \omega^e \wedge \omega^f \wedge \omega^g \wedge \omega^h \wedge \omega^i \wedge \omega^j \wedge \omega^k$$

are given by

$$\delta_b^a = \tfrac{1}{1!} \langle e_b \, \omega^a, d^1 P \rangle, \quad (61)$$
$$S_{bc}{}^a = \tfrac{1}{2!} \langle e_b \wedge e_c \, \omega^a, d^2 P \rangle, \quad (62)$$
$$\tfrac{1}{2} R_{[bcd]}{}^a = \tfrac{1}{3!} \langle e_b \wedge e_c \wedge e_d \, \omega^a, d^3 P \rangle, \quad (63)$$
$$\tfrac{1}{2} R_{[bc|f|}{}^a \, S_{de]}{}^f = \tfrac{1}{4!} \langle e_b \wedge e_c \wedge e_d \wedge e_e \, \omega^a, d^4 P \rangle, \quad (64)$$
$$\tfrac{1}{4} R_{[bc|g|}{}^a \, R_{def]}{}^g = \tfrac{1}{5!} \langle e_b \wedge e_c \wedge e_d \wedge e_e \wedge e_f \, \omega^a, d^5 P \rangle, \quad (65)$$
$$\tfrac{1}{4} R_{[bc|h|}{}^a \, R_{de|i|}{}^h \, S_{fg]}{}^i = \tfrac{1}{6!} \langle e_b \wedge e_c \wedge e_d \wedge e_e \wedge e_f \wedge e_g \, \omega^a, d^6 P \rangle, \quad (66)$$
$$\tfrac{1}{8} R_{[bc|i|}{}^a \, R_{de|j|}{}^i \, R_{fgh]}{}^j = \tfrac{1}{7!} \langle e_b \wedge e_c \wedge e_d \wedge e_e \wedge e_f \wedge e_g \wedge e_h \, \omega^a, d^7 P \rangle, \quad (67)$$
$$\tfrac{1}{8} R_{[bc|j|}{}^a \, R_{de|k|}{}^j \, R_{fg|l|}{}^k \, S_{hi]}{}^l = \tfrac{1}{8!} \langle e_b \wedge e_c \wedge e_d \wedge e_e \wedge e_f \wedge e_g \wedge e_h \wedge e_i \, \omega^a, d^8 P \rangle, \quad (68)$$
$$\tfrac{1}{16} R_{[bc|k|}{}^a \, R_{de|l|}{}^k \, R_{fg|m|}{}^l \, R_{hij]}{}^m = \tfrac{1}{9!} \langle e_b \wedge e_c \wedge e_d \wedge e_e \wedge e_f \wedge e_g \wedge e_h \wedge e_i \wedge e_j \, \omega^a, d^9 P \rangle, \quad (69)$$
$$\tfrac{1}{16} R_{[bc|l|}{}^a \, R_{de|m|}{}^l \, R_{fg|n|}{}^m \, R_{hi|o|}{}^n \, S_{jk]}{}^o = \tfrac{1}{10!} \langle e_b \wedge e_c \wedge e_d \wedge e_e \wedge e_f \wedge e_g \wedge e_h \wedge e_i \wedge e_j \wedge e_k \, \omega^a, d^{10} P \rangle \quad (70)$$

using the result that

$$\tfrac{1}{p!} \langle e_{i_1} \wedge e_{i_2} \wedge \ldots \wedge e_{i_p} \, \omega^a, d^p P \rangle = \tfrac{1}{p!} \langle e_{i_1} \wedge e_{i_2} \wedge \ldots \wedge e_{i_p} \otimes \omega^a, d^p P \rangle \quad (71)$$
$$= \tfrac{1}{p!} \langle e_{i_1} \wedge e_{i_2} \wedge \ldots \wedge e_{i_p}, \langle \omega^a, d^p P \rangle \rangle$$
$$= \tfrac{1}{p!} \langle \omega^a, \langle e_{i_1} \wedge e_{i_2} \wedge \ldots \wedge e_{i_p}, d^p P \rangle \rangle$$
$$= \begin{cases} \tfrac{1}{2^{(p-1)/2}} R_{[i_1 i_2|j_1|}{}^a \, R_{i_3 i_4|j_2|}{}^{j_1} \, R_{i_5 i_6|j_3|}{}^{j_2} \ldots R_{i_{p-4} i_{p-3}|j_{(p-3)/2}|}{}^{j_{(p-5)/2}} \, R_{i_{p-2} i_{p-1} i_p]}{}^{j_{(p-3)/2}}, & \text{if } p \text{ is odd} \\ \tfrac{1}{2^{(p-2)/2}} R_{[i_1 i_2|j_1|}{}^a \, R_{i_3 i_4|j_2|}{}^{j_1} \, R_{i_5 i_6|j_3|}{}^{j_2} \ldots R_{i_{p-3} i_{p-2}|j_{(p-2)/2}|}{}^{j_{(p-4)/2}} \, S_{i_{p-1} i_p]}{}^{j_{(p-2)/2}}, & \text{if } p \text{ is even} \end{cases}$$